\documentclass{article}
\usepackage[utf8]{inputenc}
\usepackage[english]{babel}
\usepackage{authblk}
\usepackage[T1]{fontenc}
\usepackage{amsmath,amssymb}
\usepackage{graphicx}
\usepackage[x11names]{xcolor}
\usepackage{array}
\usepackage{subcaption}
\usepackage{multirow}
\usepackage{tabularx}
\usepackage{hyperref}
\hypersetup{
    colorlinks=true,
    linkcolor=blue,
    filecolor=blue,      
    urlcolor=blue,
    citecolor=blue,
    pdfpagemode=FullScreen,
    }
\usepackage[top=2cm, bottom=2cm, left=2cm, right=2cm]{geometry}

\usepackage{ulem}

\title{\textbf{Flat-optics generation of broadband photon pairs with tunable polarization entanglement}}
\author[1,2,*]{Vitaliy Sultanov}
\author[1,2]{Tom\'{a}s Santiago-Cruz}
\author[1,2]{Maria V. Chekhova}
\affil[1]{Max-Planck Institute for the Science of Light, Staudtstr. 2, 91058 Erlangen, Germany}
\affil[2]{Friedrich-Alexander Universität  Erlangen-Nürnberg, Staudtstr. 7, 91058 Erlangen, Germany}

\affil[*]{vitaliy.sultanov@mpl.mpg.de}

\begin{document}
\maketitle


The concept of `flat optics' is quickly conquering different fields of photonics, but its implementation in quantum optics is still at infancy. In particular, polarization entanglement, which is central to quantum photonics due to the simplicity of polarization qubit encoding and control, is so far not realized on flat platforms. Meanwhile, relaxed phase matching of flat nonlinear optical sources enables enormous freedom in tailoring their polarization properties. Here we use this freedom to generate photon pairs with tunable polarization entanglement via spontaneous parametric down-conversion (SPDC) in a $400$ nm GaP film. By changing the pump polarization, we tune the polarization state of photon pairs from maximally entangled to almost disentangled, which is impossible in a single bulk source of SPDC. Polarization entanglement, in combination with the broadband frequency spectrum, results in an ultranarrow (12 fs) Hong-Ou-Mandel effect and promises extensions to hyperentanglement.

\section*{Introduction}

There is a pronounced tendency in photonics towards `flat' optics~\cite{Yu:2014}, involving ultrathin films, 2D materials, and metasurfaces. Linear and nonlinear `flat' optical elements are not only compact, integrable and efficient; they are also multifunctional, promising to replace their bulk counterparts~\cite{Chen2021}. Quantum optics is also on its way to `flat' platforms~\cite{Wang2021,Solntsev2021}. The latter are so far mainly used as hosts for single-photon sources~\cite{Toth2019} and linear converters of quantum light, both in space~\cite{Wang2018,Li2020,Li2021} and in polarization~\cite{Wang2018,Lung2020}. Meanwhile, `flat' quantum optics was so far unable to deliver polarization-entangled photons, which are ubiquitous in quantum technologies - both for quantum communications~\cite{Xu2020} and optical quantum computation~\cite{Zhong2020}. Although photon pairs have been generated in ultrathin films through spontaneous four-wave mixing~\cite{Lee2017} and spontaneous parametric down-conversion (SPDC)~\cite{Okoth2019,Santiago-Cruz2021, Grange2021, Guo2022}, it is still bulk crystals that provide polarization entanglement for flat platforms~\cite{Li2020,Lung2020}. This shortcoming hinders the development of `flat' quantum optics.

At the same time, ultrathin platforms provide unprecedented advantages in engineering photon pairs. Being free from the phase matching constraints, they can be fabricated of materials with especially large second-order susceptibility $\hat{\chi}^{(2)}$, to boost the efficiency of pair generation through SPDC. To produce polarization-entangled photons, now we use another consequence of relaxed phase matching in ultrathin SPDC sources: they enable several nonlinear interactions (type-0, type-I, type-II) simultaneously, the only restriction being the structure of the $\hat{\chi}^{(2)}$ tensor.

Here, we generate photon pairs in an ultrathin layer of gallium phosphide (GaP) whose $\hat{\chi}^{(2)}$ tensor enables SPDC of different polarization types. We find that these pairs feature nontrivial polarization properties, in particular polarization entanglement. Further, we show that by choosing the polarization of the pump we can easily tune the polarization state of the photon pairs, from maximally entangled to almost disentangled, and at the same time maintain the purity of the state. Such performance is impossible with regular linear polarization elements without introducing polarization-dependent losses in the system~\cite{Lung2020}. Moreover, in combination with the nearly unbounded frequency spectrum of photon pairs emitted from ultrathin materials~\cite{Santiago-Cruz2021}, polarization entanglement results in a remarkably narrow Hong-Ou-Mandel dip or peak, depending on the experimental conditions, which we also demonstrate in experiment. Together with the giant time-frequency~\cite{Okoth2019} and position-momentum~\cite{Okoth2020} entanglement, the observed polarization entanglement can be used to create hyperentangled photon pairs on flat platforms.

\section*{Results}

\subsection*{Tunable polarization entanglement}

\begin{figure}[ht]
\centering
\begin{subfigure}{0.5\textwidth}
\includegraphics[width=1\linewidth]{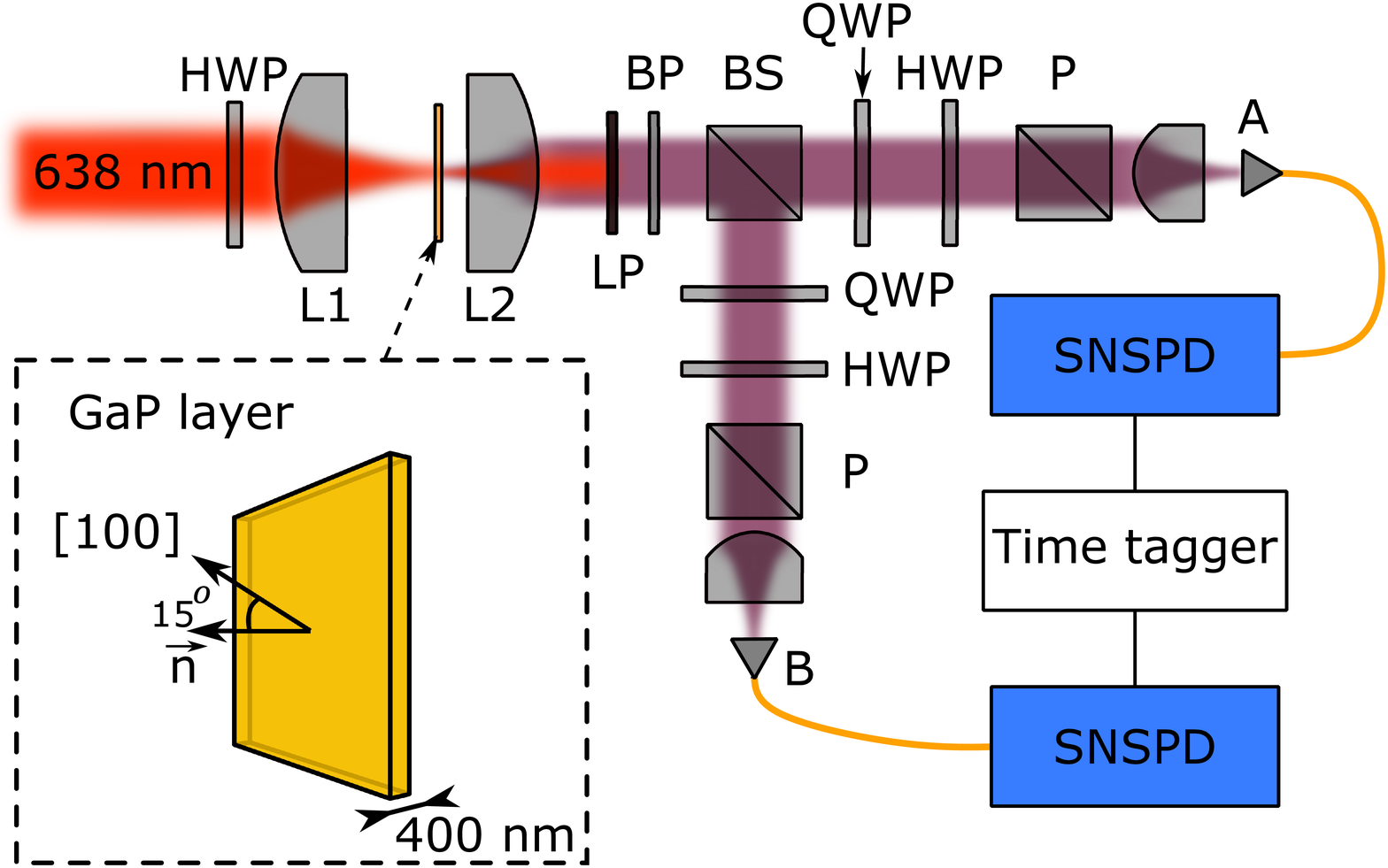}
\caption{}
\label{Setup}
\end{subfigure}
\begin{subfigure}{0.35\textwidth}
\includegraphics[width=1\linewidth]{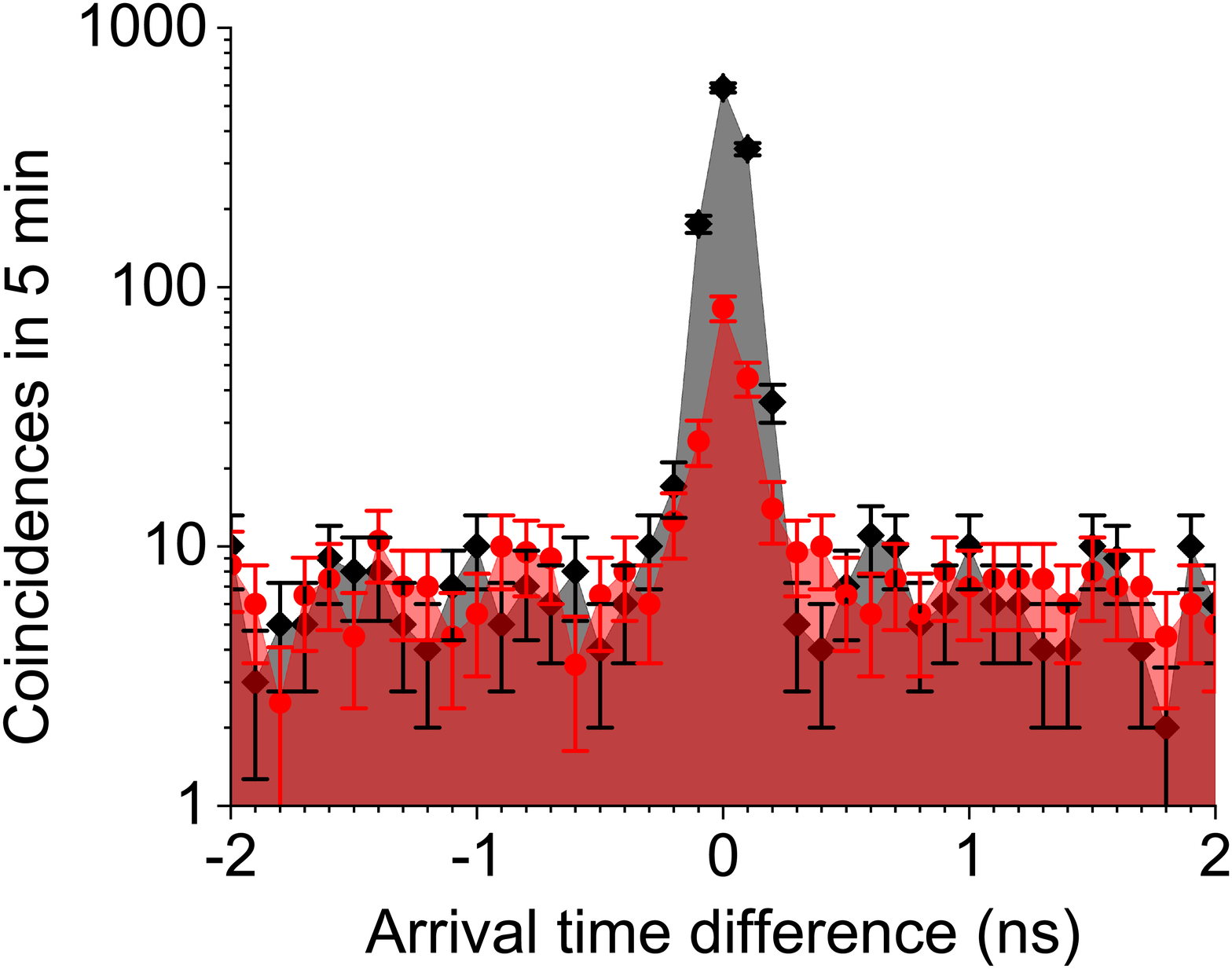}
\caption{}
\label{Histogram_example}
\end{subfigure}
\caption{\textbf{The experiment.} (\subref{Setup}) A continuous-wave pump is focused by lens L1 into a thin film of GaP, and photon pairs are collected by lens L2 and filtered from the pump by long-pass filters LP and band-pass filter BP. Non-polarizing beam splitter BS sends the photons into arms A and B, each containing a quarter-wave plate (QWP), a half-wave plate (HWP), a polarizer (P), and a superconducting nanowire single-photon detector (SNSPD). A time tagger builds a histogram of arrival time differences (\subref{Histogram_example}). The histogram of is plotted in the logarithmic scale for the case of the horizontally (black) and vertically (red) polarized pump.}

\label{Setup-spectrum}
\end{figure}

We generate photon pairs via SPDC in a 400 nm-thick film of GaP, pumped by a continuous-wave laser at $638$ nm (Fig.~\ref{Setup}), whose linear polarization can be rotated by a half-wave plate (HWP). 
The GaP film is fabricated on the fused silica substrate with its normal at $15^\circ$ to the [100] crystalline direction, so that the pump field, in the general case, has projections on all crystallographic axes $x,y,z$. The efficiency of SPDC scales as $|\hat{\chi}^{(2)}\cdot\vec{e}_s\cdot\vec{e}_i\cdot\vec{e}_p|^2$, where $\vec{e}_{s,i,p}$ are the unit polarization vectors of the signal, idler, and pump radiation, respectively~\cite{ChekhovaPolar}. Because of the relaxed phase matching condition, all  nonzero elements of the $\hat{\chi}^{(2)}$ tensor of GaP contribute to SPDC; in particular, there is emission of pairs for any pump polarization (see the supporting information (SI) for details).

After collecting the SPDC radiation and filtering it from the pump, we send it to the Hanbury Brown - Twiss setup~\cite{HBT1956} formed by non-polarizing beam splitter BS and superconducting nanowire single-photon detectors (SNSPDs) in two output ports A and B. A time tagger receives photo-detection pulses from both detectors and builds the histogram of their time differences (Fig. \ref{Histogram_example}), where the central peak clearly shows the simultaneity of photon arrivals in both arms. The peak is observed for both vertically (V) and horizontally (H) polarized pump (red and black points). Events forming this peak, called `two-photon coincidences', indicate the detections of photon pairs. The background of the histogram shows accidental two-photon coincidences, and its level is determined by the product of the two detectors' count rates. The same level of the background for the significantly different number of coincidences shows that accidental coincidences are almost entirely caused by the photoluminiscence of the GaP film. In all measurements described below, this background is subtracted.

We see that, in contrast to SPDC in bulk crystals, photon pairs are emitted for both V-polarized and H-polarized pump, although at a higher rate in the latter case (Fig.\ref{Histogram_example}). Even more strikingly, the two cases result in different polarization states of the photon pairs.

A general polarization state of a photon pair in a single frequency and wavevector mode can be written as a qutrit: a superposition of vertically, horizontally, and orthogonally polarized photon pairs~\cite{Burlakov1999},
\begin{equation}
    |\Psi\rangle = C_1|2\rangle_H|0\rangle_V+C_2|1\rangle_H|1\rangle_V+C_3|0\rangle_H|2\rangle_V,
    \label{qutrit}
\end{equation}
where $|N\rangle_{H, V}$ are the Fock states with $N$ photons polarized horizontally or vertically, respectively, and $C_i$ are the complex amplitudes satisfying $\sum_{i=1}^3|C_i|^2=1$.

For state~(\ref{qutrit}), the degree of polarization entanglement is given by the concurrence~\cite{Fedorov2014} 
\begin{equation}
    C = \left| 2C_1C_3 - C_2^2\right|,
    \label{concurrence}
\end{equation}
taking the minimal value $C=0$ for a pair of co-polarized photons and the maximal value $C=1$ for a pair of orthogonally polarized photons. Accordingly, the Schmidt number $K=2/(2-C^2)$, another measure of entanglement, takes in these cases the values $1$ and $2$, respectively~\cite{Fedorov2014}.

To analyse the polarization state of the pairs generated in our experiment, we apply two-photon polarization tomography~\cite{Burlakov2003}, which enables the reconstruction of state (\ref{qutrit}) or, for a mixed state, of the corresponding density matrix $\rho$. The procedure is similar to the Stokes measurements on single photons~\cite{ChekhovaPolar}, but involves coincidence registration. In two-photon polarization tomography, we independently select different polarization states in each  beam splitter output port and measure the rate of coincidences. The polarization is selected by a quarter-wave plate (QWP) and a half-wave plate (HWP), followed by a polarizer (P), see Fig.~\ref{Setup}. This combination, further called a polarization analyzer, can filter photons in an arbitrary polarization state. The combination of two such analyzers in two arms filters an arbitrary two-photon polarization state. By measuring the rate of coincidences for nine such combinations, we fully characterize the two-photon polarization state before the beam splitter (see ``Methods'').

\begin{figure}[ht]
\centering
\begin{tabular}[c]{cc}
\begin{subfigure}{0.223\textwidth}
\includegraphics[width=1\linewidth]{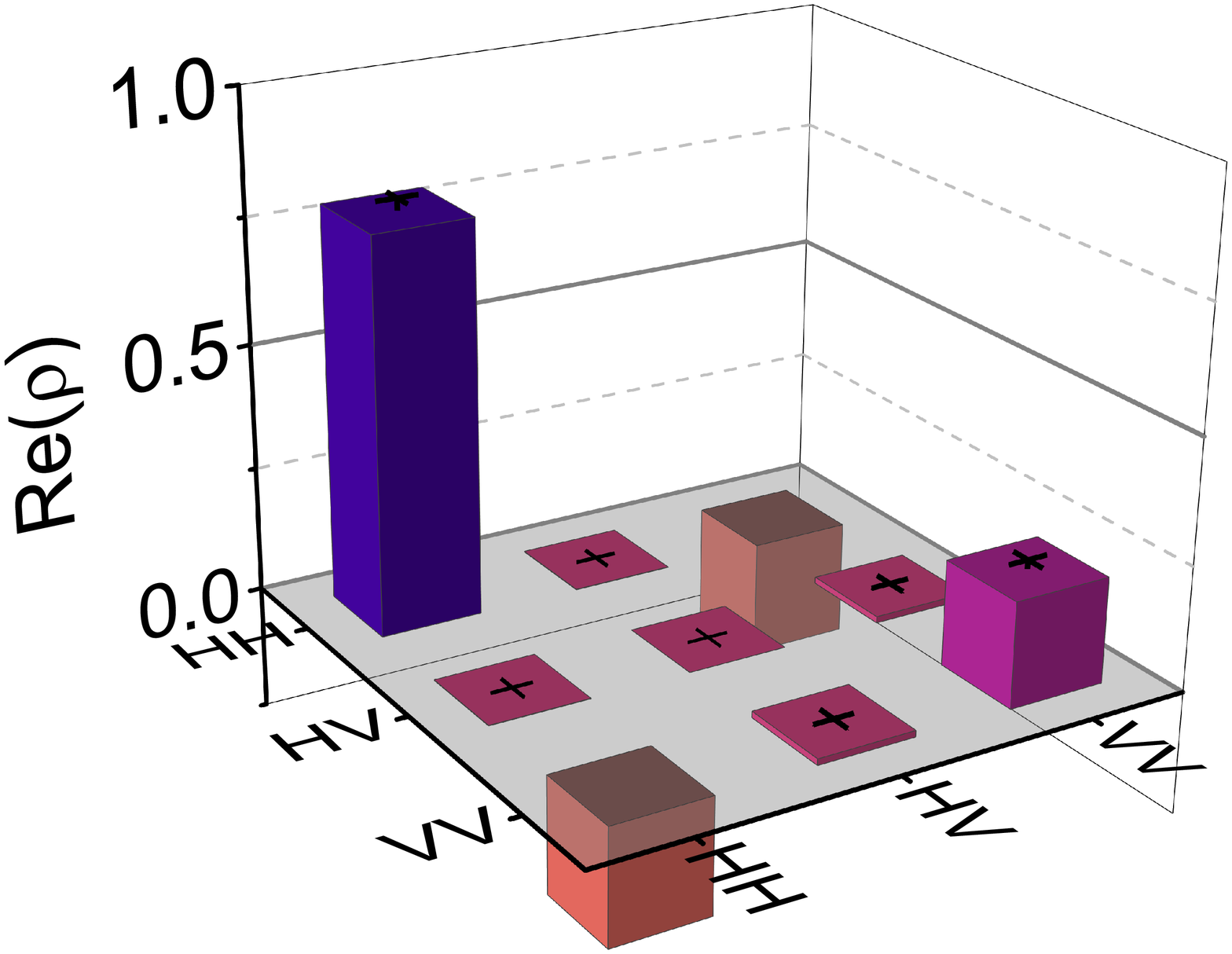}
\caption{}
\label{GaP_H_pump_real}
\end{subfigure}&
\begin{subfigure}{0.257\textwidth}
\includegraphics[width=1\linewidth]{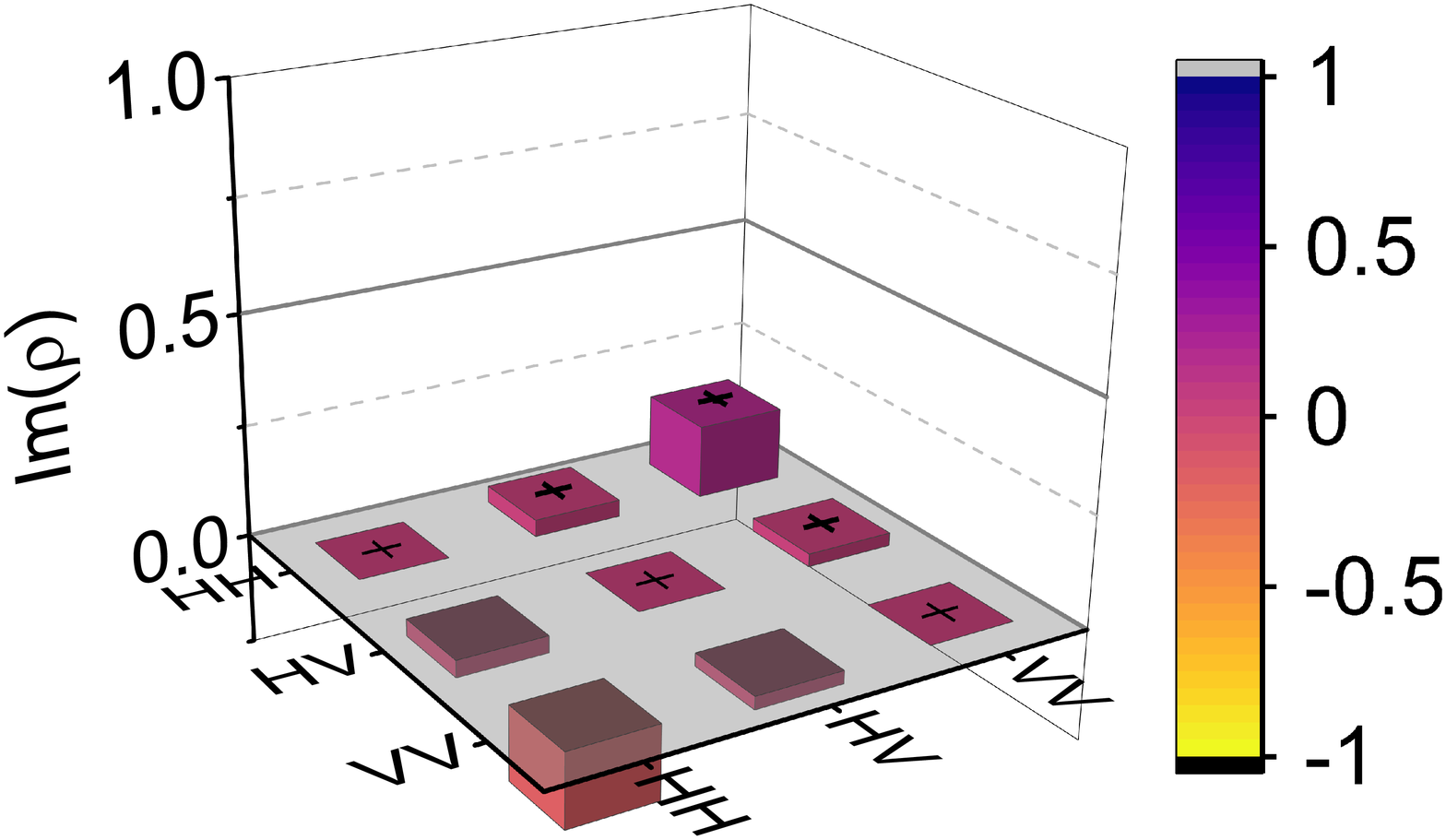}
\caption{}
\label{GaP_H_pump_imag}
\end{subfigure}\\
\begin{subfigure}{0.223\textwidth}
\includegraphics[width=1\linewidth]{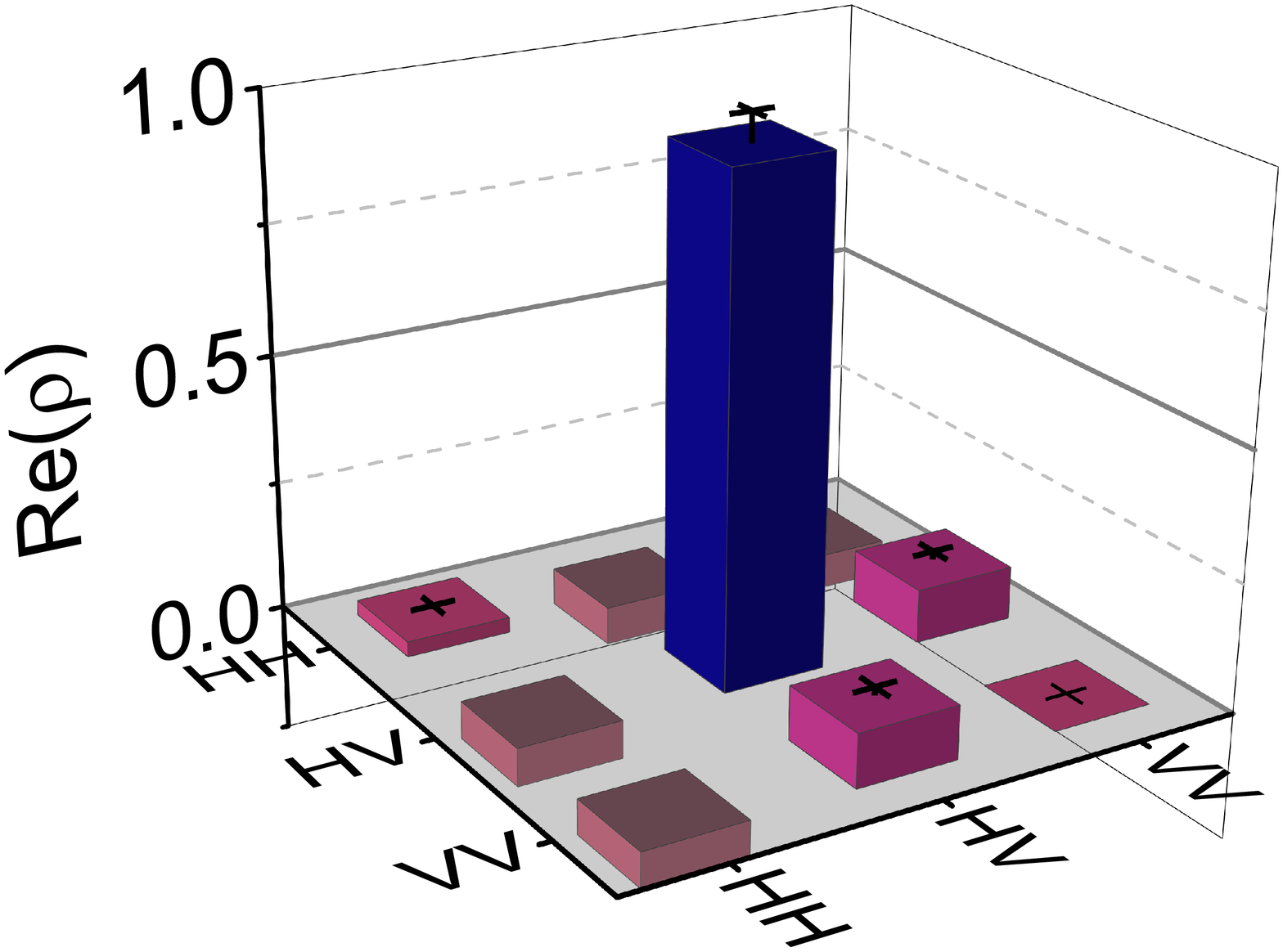}
\caption{}
\label{GaP_V_pump_real}
\end{subfigure}&
\begin{subfigure}{0.257\textwidth}
\includegraphics[width=1\linewidth]{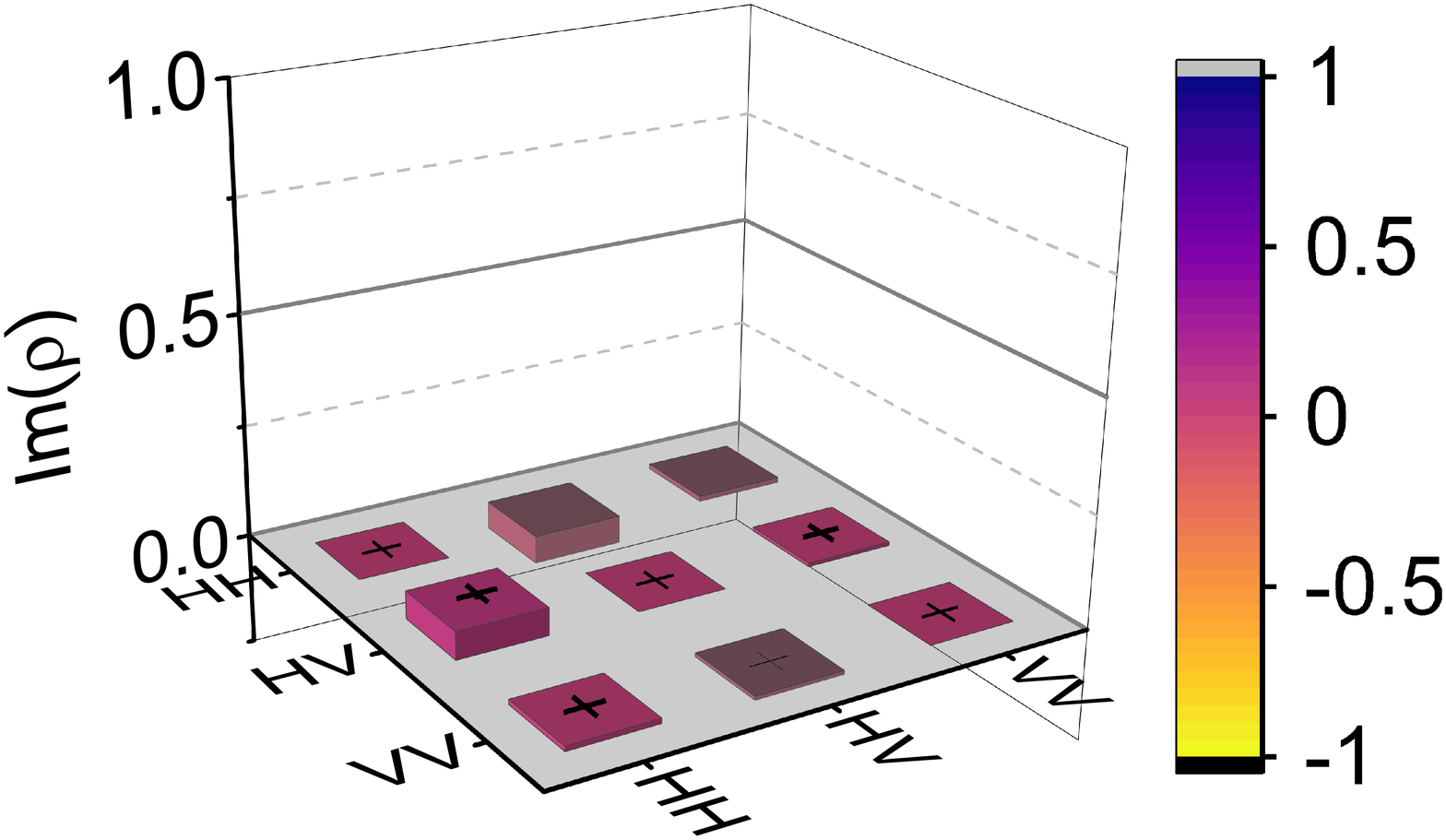}
\caption{}
\label{GaP_V_pump_imag}
\end{subfigure}\\
\end{tabular}
\begin{subfigure}{0.4\textwidth}
\includegraphics[width=1\linewidth]{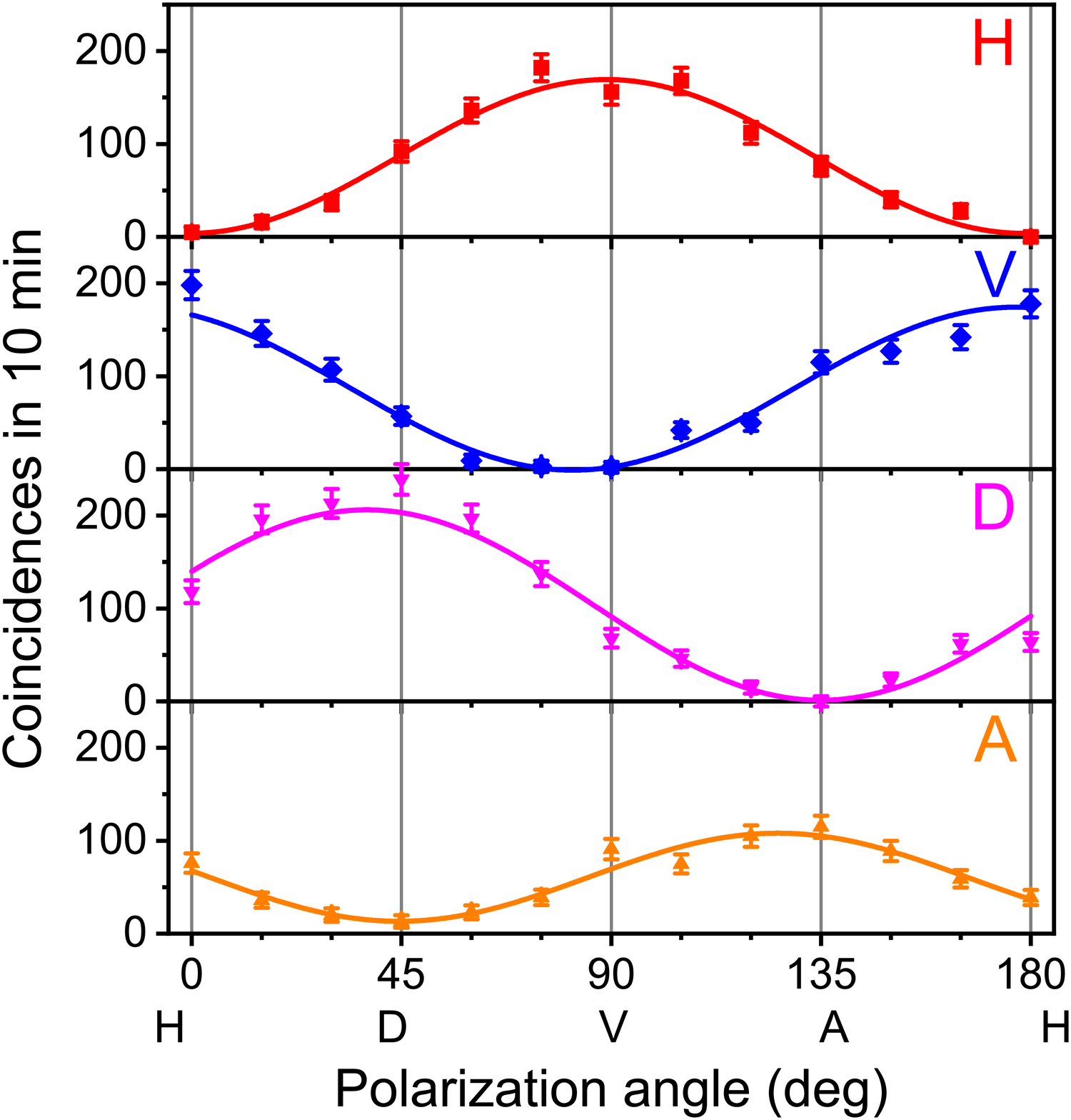}
\caption{}
\label{Polarization_entanglement}
\end{subfigure}
\caption{\textbf{Two-photon state reconstruction and polarization entanglement.} Panels \subref{GaP_H_pump_real} (\subref{GaP_V_pump_real}) and \subref{GaP_H_pump_imag} (\subref{GaP_V_pump_imag}) show the real and imaginary parts of the density matrix $\rho$ of the photon pairs generated by the H- (V-) polarized pump. (\subref{Polarization_entanglement}) The number of coincidences versus the linear polarization angle selected in arm A for the case of V-polarized pump (orthogonally polarized photons). In arm B, we select the basic linearly polarized states: horizontal (H), vertical (V), diagonal (D), anti-diagonal (A). The experimental points are fitted with the theoretical dependence for polarization-entangled photons (solid lines). The visibility reaches 96\%, indicating a high degree of polarization entanglement.}
\label{SPDC_PT_PE}
\end{figure}

We run this procedure for the states produced with H - and V - polarized pump, using a $50$ nm FWHM band-pass filter centered at 1275 nm to suppress the photoluminescence.
The reconstructed density matrices of photon pairs generated by the H - and V - polarized pump are shown in Fig. \ref{GaP_H_pump_real}-\subref{GaP_H_pump_imag}, and \ref{GaP_V_pump_real}-\subref{GaP_V_pump_imag}, respectively. 

For the H - polarized pump, the two-photon polarization state is a superposition of the horizontally and vertically polarized pairs with different weights: $|C_1|^2 = 0.79 \pm 0.02$, $|C_2|^2=0$, and $|C_3|^2 = 0.21 \pm 0.02$. Since $|C_1|^2\gg|C_3|^2$, two photons within a pair are almost co-polarized horizontally. In contrast, pairs generated with the V - polarized pump contain orthogonally (HV) polarized photons: $|C_1|^2 = 0.03 \pm 0.01$, $|C_2|^2 = 0.97 \pm 0.06$, and $|C_3|^2 = 0$. These results demonstrate a vast tunability of the two-photon polarization state, which is not achievable with a single bulk SPDC source and requires additional linear elements~\cite{White1999}. 

Moreover, by changing the polarization of the pump we also tune the degree of polarization entanglement. In the case of an H - polarized pump, Eq.~(\ref{concurrence}) yields the concurrence $C=0.4\pm0.03$ and the Schmidt number $K = 1.09 \pm 0.01$, i.e., photons within a pair are almost disentangled. For the HV photon pairs, obtained from a V - polarized pump, the degree of entanglement is maximal, $C=0.98\pm0.02$ and $K = 1.9 \pm 0.1$. Notably, this tunability is impossible with standard linear polarization elements, which maintain the degree of entanglement and can modify it only through lossy transformations~\cite{Lung2020}.

\subsection*{Bell inequality violation}

We further demonstrate the polarization entanglement of photon pairs generated with the V - polarized pump by measuring the number of coincidences for different orientations of the analyzers. The analyzer in arm B is fixed at one of the basic polarization states, horizontal (H), vertical (V), diagonal (D), or anti-diagonal (A). The analyzer in arm A is then selecting linear polarization, rotated from $0^\circ$ to $180^\circ$. The results are shown in Fig.~\ref{Polarization_entanglement}. A visibility of 96\% witnesses strong polarization entanglement. 

By splitting photon pairs generated with the V - polarized pump with the non-polarizing beam splitter, we create the Bell state $\left|\Psi^{(+)}\right>$ (see SI) and use it to test the Bell inequality in the Clauser-Horne-Shimony-Holt form~\cite{CHSH1969},
\begin{equation}
    F\equiv\frac{1}{2}|\langle ab+a'b+ab'-a'b'\rangle|\le 1.
    \label{Bell}
\end{equation}
We chose the binary variables $a, \; a',\; b,\; b' = \pm 1$ for photons in arms A and B as their Stokes operators $S_1^A$, $-S_2^A$, $\frac{1}{\sqrt{2}}(S_1^B+S_2^B)$ and $\frac{1}{\sqrt{2}}(S_1^B-S_2^B)$, respectively~\cite{Klyshko1998}. By measuring simultaneously the Stokes observables of photons A, B in the two output ports of the beam splitter and calculating the left-hand side of inequality (\ref{Bell}), we obtain $F = 1.36 \pm 0.07$, which violates the Bell inequality by five standard deviations.

 Remarkably, generation of polarization-entangled photon pairs with high purity does not require any additional elements erasing the distinguishability between pairs of different polarization states, as is the case in bulk phase-matched sources~\cite{Kwiat1995}.  For orthogonally polarized photons, we obtain the value of the purity  $Tr(\rho^2) = 1.0 \pm 0.1$. The indistinguishability stems from the isotropic linear optical properties of GaP. A similar feature is expected in other nonlinear crystals with zinc-blende structures like GaAs or InAs~\cite{Boyd}.

\begin{figure}[ht]
\centering
\begin{tabular}[c]{cc}
\begin{subfigure}{0.35\textwidth}
\includegraphics[width=1\linewidth]{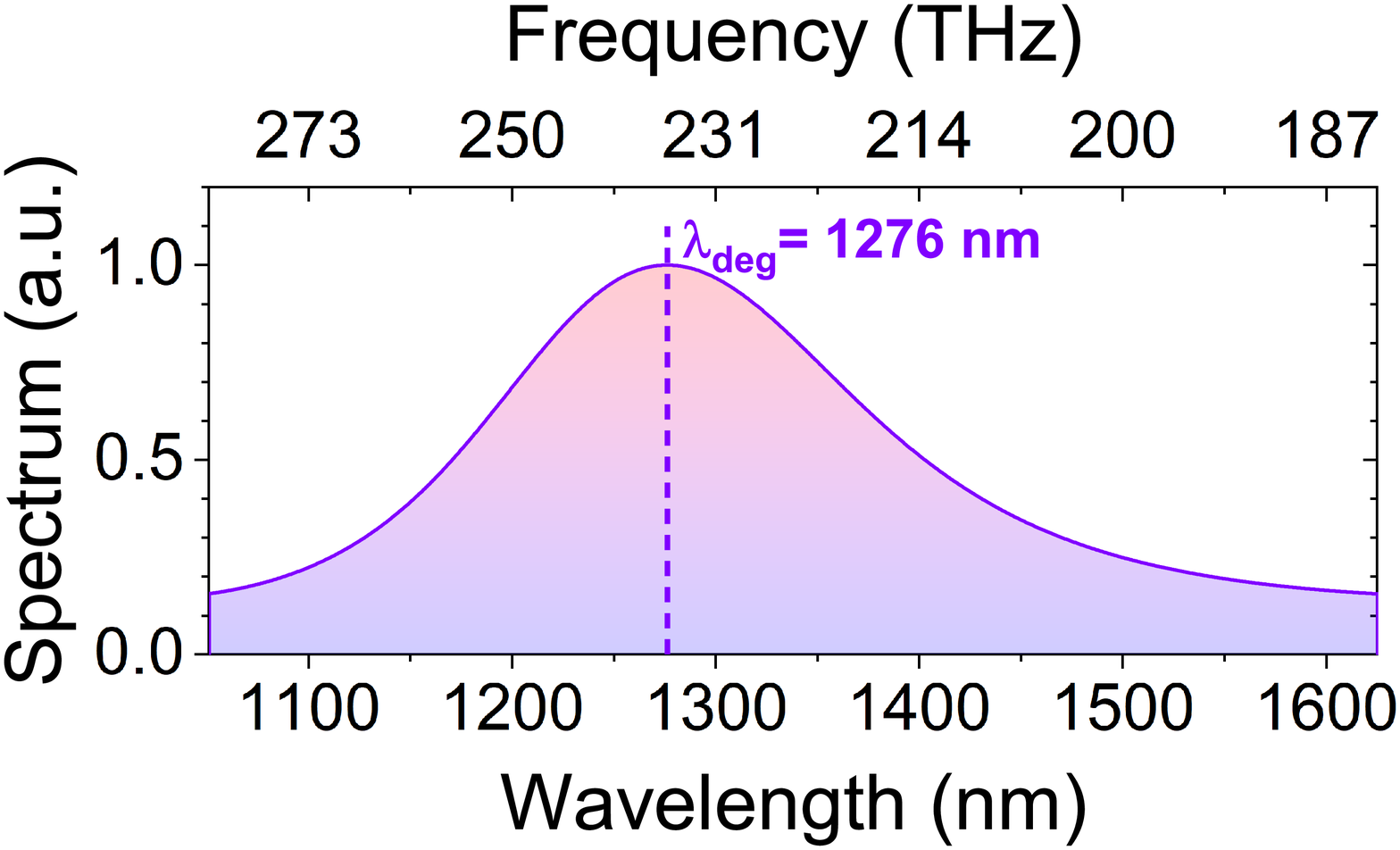}
\caption{}
\label{Spectrum}
\end{subfigure}\\
\begin{subfigure}{0.35\textwidth}
\includegraphics[width=1\linewidth]{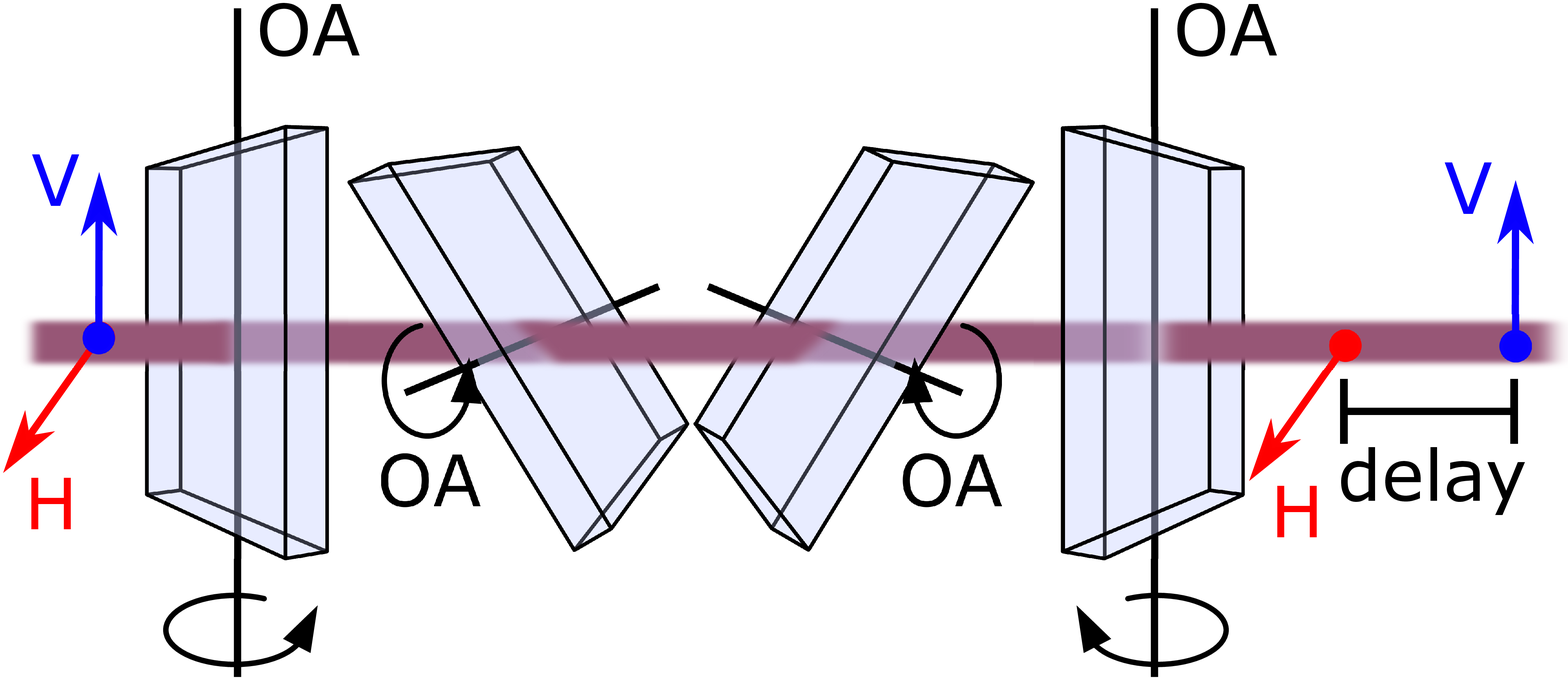}
\caption{}
\label{Calcite_delay}
\end{subfigure}
\end{tabular}
\begin{subfigure}{0.6\textwidth}
\vspace{1cm}
\includegraphics[width=1\linewidth]{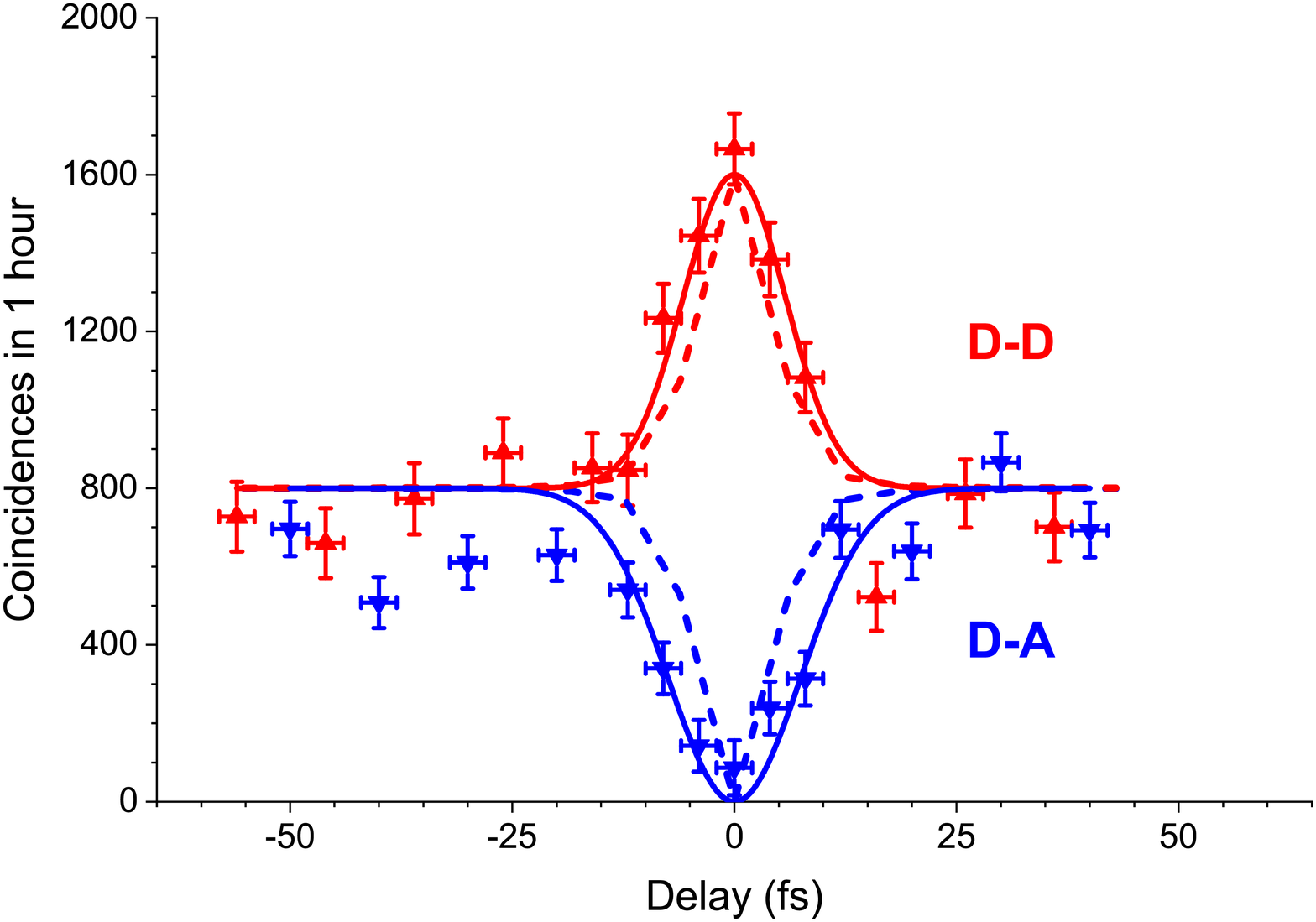}
\caption{}
\label{HOM}
\end{subfigure}
\caption{\textbf{Broadband polarization entanglement.} (\textbf{\subref{Spectrum}}) Calculated spectrum of the photon pairs, with the width mainly limited by the Fabry-Perot effect in the film. (\subref{Calcite_delay}) The delay line for the HOM experiment consists of 4 calcite plates, 5 mm thick each. The birefringence of calcite introduces a delay between orthogonally polarized photons, which is varied by tilting the plates around their optic axes (OA). (\textbf{\subref{HOM}}) The HOM dip (blue) and peak (red), measured (points) and calculated based on the Fourier transform of the spectrum in panel (\textbf{\subref{Spectrum}}) (dashed lines). Solid lines show the Gaussian fit of the experimental data. The experimental width of the dip (peak) is $15\pm2$ fs ($12\pm2$ fs), which is somewhat broader than that predicted by the calculated spectrum ($10$ fs).}
\label{Spectrum_HOM}
\end{figure}

\subsection*{Hong-Ou-Mandel effect}

As we see in Fig.~\ref{Polarization_entanglement}, H-V photon pairs generated by the V - polarized pump result in no coincidences if the polarization analyzers select the D-A polarization states. This is the polarization version of the Hong-Ou-Mandel (HOM) effect~\cite{HOM1987, Rubin1994} and another manifestation of polarization entanglement. Destructive quantum interference leads to the absence of coincidences when the D-A polarization state is detected. In contrast, when the analyzers are both oriented diagonally (D-D) or anti-diagonally (A-A), we observe the maximal rate of coincidences due to the constructive interference. This behavior is similar to the `standard' version of the HOM effect: two photons arriving from different input ports of a beam splitter (here, from orthogonal polarization modes H,V) are both directed into one of the output ports (here, into polarization mode D or A).

By introducing a time delay between the photons of a pair, one observes a dip in the case of destructive interference and a peak in the case of constructive interference. The width of the dip or peak is determined by the spectrum of the photon pairs~\cite{Burlakov2001}. In our case, the spectrum is ultrabroad due to the tiny thickness of the sample. Figure~\ref{Spectrum} shows the calculated spectrum, with the etalon effect taken into account~\cite{Kitaeva2004, URen2010}. Despite the reduction due to the etalon effect~\cite{Santiago-Cruz2021}, the spectral width is still as large as $50$ THz, which should lead to the HOM dip and peak of width about $10$ fs (dashed lines in Fig.~\ref{HOM}).

In the experiment, we introduce a variable time delay between orthogonally (H-V) polarized photons before the non-polarizing beam splitter by means of four 5 mm birefringent calcite plates (Fig. \ref{Calcite_delay}, Methods).


The results of the HOM experiment are shown in Fig.~\ref{HOM}. The blue (red) triangles show the number of coincidences versus the delay between the photons for the D-A (D-D) selected polarization state. The solid curves are theoretical dependences calculated using the theoretical spectrum in Fig.~\ref{Spectrum}. As expected, we see a dip of coincidences at zero delay when the analyzers select orthogonal polarization states, D-A, and a peak when they select the same polarization states, D-D. At large time delays, quantum interference disappears because single-photon wavepackets do not overlap in time. The widths of the dip and the peak are $15\pm2$ fs and $12\pm2$ fs, respectively, which is somewhat larger than the width predicted from the spectrum in Fig.~\ref{Spectrum} ($10$ fs). This is because the two-photon spectrum is additionally narrowed by the spectrally dependent efficiency of the detectors.

\section*{Discussion}

We show that the relaxed phase matching for SPDC in an ultrathin nonlinear film leads to the unprecedented polarization tunability of the produced photon pairs. Taking advantage of the $\hat{\chi}^{(2)}$ tensor form, by simply adjusting the pump polarization, we drastically tune the two-photon polarization state and change it from maximally polarization-entangled to almost disentangled. This tunability provides enormous freedom in photon pair polarization engineering, impossible in conventional systems without introducing losses. With such a system, one can change the degree of polarization entanglement on demand in a very easy way, which is useful for various applications of quantum technologies. In particular, two-photon states with variable degree of entanglement are of interest for quantum key distribution~\cite{Xue2001}, teleportation~\cite{Modlawska2008} and Bell inequality tests~\cite{Giustina2015}.

To verify the high degree of polarization entanglement of one of the states, we experimentally violate the Bell inequality by five standard deviations. This is the first observation of polarization-entangled photons from ultrathin nonlinear films. The polarization-entangled state is of high purity and its preparation does not require linear optical polarization compensators, unavoidable with bulk SPDC sources.

Polarization entanglement comes in combination with an ultrabroad frequency spectrum, an extremely high degree of time/frequency entanglement~\cite{Okoth2019} and ultra-narrow time correlations. Accordingly, we observe very narrow ($12-15$ fs) polarization HOM dip and peak for polarization-entangled photon pairs. This feature can be used for polarization-sensitive quantum-optical coherence tomography~\cite{Booth2004}. We note that the width of the HOM dip and peak can be further reduced by improving the detection setup.

The combination of polarization with other degrees of freedom enables achieving hyperentanglement, which allows superdense coding, considerably increasing the information capacity of two-photon states~\cite{Graham2015}. So far, SPDC in ultrathin films resulted in photon pairs with huge entanglement in frequency~\cite{Okoth2019} and momentum~\cite{Okoth2020}. Now, we complement this set with the polarization degree of freedom. Further extension can be done by considering entanglement in orbital-angular momentum (OAM) of photons~\cite{Xu2021}.

Finally, the same flexibility of polarization properties can be used in  metasurface sources of SPDC, where the pair production rate is enhanced due to geometric resonances~\cite{Santiago-Cruz2021_Nano}. The polarization state of the photon pairs will then depend not only on the form of the nonlinear tensor, but also on the resonant field distribution in the meta-atoms.

\section*{Methods}

\textbf{Sample.} As a source of photon pairs, we use a $400$ nm film of GaP, which is a semiconductor with a zinc-blende symmetry. The film is fabricated on  a $4$ $\mu$m layer of SiO$_2$, which, in turn, is deposited on a $150$ $\mu$m sapphire substrate. The fabrication procedure is described in Ref.~\cite{Anthur2021}. The non-zero $\hat{\chi}^{(2)}$ components of GaP are $\chi_{xyz} = \chi_{xzy} = \chi_{yzx} = \chi_{yxz} = \chi_{zyx} = \chi_{zxy}\approx100$ pm/V~\cite{Boyd,Anthur2021}. To exploit these components, the sample was grown with the orientation as shown in the inset of Fig.~\ref{Setup}.

\noindent\textbf{Photon pair generation and detection.} The pump (60 mW continuous-wave diode laser centered at 638 nm) is focused into the sample with lens L1 (Fig. \ref{Setup}), covering NA=$0.05$. Another lens (L2, NA$=0.16$) collects the generated photon pairs. We cut off the pump with four long-pass filters (LP) with the cut-on wavelengths $850$ nm and $1000$ nm. For polarization measurements, we additionally select frequency-degenerate photon pairs with a band-pass filter (BP) centered at 1275 nm with 50 nm full width at half maximum. A non-polarizing beam splitter (BS) splits photon pairs into arms A and B, each of them containing a quarter-wave plate (QWP), a half-wave plate (HWP), a polarizer (P), and a superconducting nanowire single-photon detector (SNSPD).

\noindent\textbf{Two-photon polarization reconstruction.} To reconstruct the single-mode two-photon polarization state, we use polarization tomography of single-mode photon pairs~\cite{Burlakov2003}. Since the density matrix of the two-photon polarization state has a dimension of $n=3$, $n^2=9$ measurements (including one extra measurement for normalization) of the coincidence count rate are required to fully reconstruct the state and verify its purity. The full protocol of the measurements  for nine combinations of the analyzer settings in arms A,B can be found in SI. In the more general case of photon pairs generated in two spatial modes, the dimension of the reduced density matrix is 4, and, as a consequence, 16 measurements are required for quantum tomography~\cite{White1999}. 

\noindent\textbf{Time delays in Hong-Ou-Mandel effect.} 
To introduce the time delay between orthogonally polarized photons, we use four $5$ mm thick calcite crystal plates (Fig. \ref{Calcite_delay}), with the optic axes vertical for two and horizontal for the other two plates. Tilting the plates around their optic axes changes their effective thickness. Due to the birefringence of calcite, the optical path for the horizontally and vertically polarized photons is different, which leads to a delay between them depending on the tilt angle of a plate. The zero delay is achieved when two orthogonally oriented crystal plates are tilted by the same angle. To compensate for the transverse displacement of the beam, we put a mirrored scheme with another two plates. By rotating the inner or the outer pair of plates we are able to achieve either negative or positive delay between two orthogonally polarized photons.

\section*{Data availability}

Any data that support the findings of this study are available from the corresponding author upon reasonable request.

\section*{Aknowledgements}

We thank H. Zhang, A. P. Anthur and L. A. Krivitsky for providing the sample and A. V. Rasputnyi for the help at the initial stage of the experiment. V.S. and T.S.C. are part of the Max Planck School of Photonics supported by BMBF, Max Planck Society, and
Fraunhofer Society. This work was funded by the Deutsche Forschungsgemeinschaft (DFG, German Research Foundation) – Project-ID 429529648 – TRR 306 QuCoLiMa (“Quantum Cooperativity of Light and Matter’’).

\section*{Author contribution}

V.S. built the experimental setup, carried out the experiments, analyzed and interpreted the experimental data. T.S.C. significantly contributed to the experiment and the writing of the manuscript. M. V. C. conceived the idea of the experiment, supervised the project, and significantly revised the manuscript.

\section*{Competing interest}

The authors declare no competing interests.

\bibliographystyle{UNSRT}

\end{document}